\input harvmac
\noblackbox

\lref\Yi{ P.~Yi,
``Anomaly of (2,0) theories,''
Phys.\ Rev.\ D {\bf 64} (2001) 106006
[arXiv:hep-th/0106165].}

\lref\howe{P.~S.~Howe, N.~D.~Lambert and P.~C.~West,
``The self-dual string soliton,''
Nucl.\ Phys.\ B {\bf 515} (1998) 203
[arXiv:hep-th/9709014]. }

\lref\Callan{
C.~G.~.~Callan and J.~M.~Maldacena,
``Brane dynamics from the Born-Infeld action,''
Nucl.\ Phys.\ B {\bf 513}, 198 (1998)
[arXiv:hep-th/9708147].
}

\lref\Gibbons{
G.~W.~Gibbons,
``Born-Infeld particles and Dirichlet p-branes,''
Nucl.\ Phys.\ B {\bf 514}, 603 (1998)
[arXiv:hep-th/9709027].
}

\lref\brax{P.~Brax and J.~Mourad,
``Open supermembranes coupled to M-theory five-branes,''
Phys.\ Lett.\ B {\bf 416} (1998) 295
[arXiv:hep-th/9707246]. }

\lref\intril{K.~A.~Intriligator,
 ``Anomaly matching and a Hopf-Wess-Zumino term in 6d, N = (2,0) field
theories,''
Nucl.\ Phys.\ B {\bf 581} (2000) 257
[arXiv:hep-th/0001205]. }

\lref\ganor{O.~Ganor and L.~Motl,
``Equations of the (2,0) theory and knitted fivebranes,'' JHEP {\bf
9805} (1998) 009 [arXiv:hep-th/9803108]. }

\lref\harv{J.~A.~Harvey, R.~Minasian and G.~W.~Moore,
``Non-abelian tensor-multiplet anomalies,'' JHEP {\bf 9809} (1998) 004
[arXiv:hep-th/9808060].}

\lref\freed{D.~Freed, J.~A.~Harvey, R.~Minasian and G.~W.~Moore,
``Gravitational anomaly cancellation for M-theory fivebranes,'' Adv.\
Theor.\ Math.\ Phys.\  {\bf 2} (1998) 601 [arXiv:hep-th/9803205].}

\lref\michael{M.~B.~Green, J.~A.~Harvey and G.~W.~Moore,
``I-brane inflow and anomalous couplings on D-branes,'' Class.\
Quant.\ Grav.\  {\bf 14} (1997) 47 [arXiv:hep-th/9605033].}

\lref\henningson{M.~Henningson and K.~Skenderis,
``The holographic Weyl anomaly,'' JHEP {\bf 9807} (1998) 023
[arXiv:hep-th/9806087].}

\lref\Klebanov{I.~R.~Klebanov,
``World-volume approach to absorption by non-dilatonic branes,''
Nucl.\ Phys.\ B {\bf 496} (1997) 231 [arXiv:hep-th/9702076].}

\lref\wittfive{E.~Witten,
``Five-brane effective action in M-theory,'' J.\ Geom.\ Phys.\  {\bf
22} (1997) 103 [arXiv:hep-th/9610234].}

\lref\berman{D.~S.~Berman,
``Aspects of M-5 brane world volume dynamics,'' Phys.\ Lett.\ B {\bf
572} (2003) 101 [arXiv:hep-th/0307040].
D.~S.~Berman and P.~Sundell, ``AdS(3) OM theory and the self-dual
string or membranes ending on the
Phys.\ Lett.\ B {\bf 529} (2002) 171 [arXiv:hep-th/0105288].
}

\lref\sevrin{X.~Bekaert, M.~Henneaux and A.~Sevrin,
``Chiral forms and their deformations,'' Commun.\ Math.\ Phys.\  {\bf
224} (2001) 683 [arXiv:hep-th/0004049].}

\lref\fiveeom{
P.~S.~Howe, E.~Sezgin and P.~C.~West,
``Covariant field equations of the M-theory five-brane,''
Phys.\ Lett.\ B {\bf 399} (1997) 49
[arXiv:hep-th/9702008].}

\lref\Minasian{
R.~Minasian and G.~W.~Moore,
JHEP {\bf 9711}, 002 (1997)
[arXiv:hep-th/9710230].
}

\lref\Cheung{
Y.~K.~Cheung and Z.~Yin,
Nucl.\ Phys.\ B {\bf 517}, 69 (1998)
[arXiv:hep-th/9710206].
}

\lref\Henningson{
M.~Henningson,
``Self-dual strings in six dimensions: Anomalies, the ADE classification, and
the world-sheet WZW-model,''
arXiv:hep-th/0405056.
}

\lref\bott{R.~ Bott, L.~ W.~ Tu, ``Differential Forms in Algebraic Topology'', Springer Verlag 1982.}

\lref\lechner{
K.~Lechner, ``Intersecting M2- and M5-branes,'' arXiv:hep-th/0402078.
}

\lref\pope{
S.~Deser, A.~Gomberoff, M.~Henneaux and C.~Teitelboim,
``Duality, self-duality, sources and charge quantization in abelian  N-form
theories,''
Phys.\ Lett.\ B {\bf 400} (1997) 80
[arXiv:hep-th/9702184].
S.~Deser, A.~Gomberoff, M.~Henneaux and C.~Teitelboim,
``p-brane dyons and electric-magnetic duality,''
Nucl.\ Phys.\ B {\bf 520} (1998) 179
[arXiv:hep-th/9712189].
}

\lref\Morales{
J.~F.~Morales, C.~A.~Scrucca and M.~Serone,
``Anomalous couplings for D-branes and O-planes,''
Nucl.\ Phys.\ B {\bf 552}, 291 (1999)
[arXiv:hep-th/9812071].
B.~J.~Stefanski,
``Gravitational couplings of D-branes and O-planes,''
Nucl.\ Phys.\ B {\bf 548} (1999) 275
[arXiv:hep-th/9812088].
}

\lref\CrapsII{
B.~Craps and F.~Roose,
``(Non-)anomalous D-brane and O-plane couplings: The normal bundle,''
Phys.\ Lett.\ B {\bf 450}, 358 (1999)
[arXiv:hep-th/9812149].
}

\lref\CrapsI{
B.~Craps and F.~Roose,
``Anomalous D-brane and orientifold couplings from the boundary state,''
Phys.\ Lett.\ B {\bf 445}, 150 (1998)
[arXiv:hep-th/9808074].
}

\lref\Bershadsky{
M.~Bershadsky, C.~Vafa and V.~Sadov,
``D-Branes and Topological Field Theories,''
Nucl.\ Phys.\ B {\bf 463}, 420 (1996)
[arXiv:hep-th/9511222].
}

\lref\thompson{
C.~Boulahouache and G.~Thompson,
``One loop effects in various dimensions and D-branes,''
Int.\ J.\ Mod.\ Phys.\ A {\bf 13} (1998) 5409
[arXiv:hep-th/9801083].
}

\lref\tseytlin{ 
A.~A.~Tseytlin and K.~Zarembo,
``Magnetic interactions of D-branes and Wess-Zumino terms in super  Yang-Mills
Phys.\ Lett.\ B {\bf 474} (2000) 95
[arXiv:hep-th/9911246].
}

\lref\gsanom{
M.~B.~Green and J.~H.~Schwarz,
``Anomaly Cancellation In Supersymmetric D=10 Gauge Theory And Superstring
Theory,''
Phys.\ Lett.\ B {\bf 149}, 117 (1984).
}

\lref\beckers{
K.~Becker and M.~Becker,
``Fivebrane gravitational anomalies,''
Nucl.\ Phys.\ B {\bf 577}, 156 (2000)
[arXiv:hep-th/9911138].
}

\lref\kaplan{
D.~M.~Kaplan and J.~Michelson,
``Zero Modes for the D=11 Membrane and Five-Brane,''
Phys.\ Rev.\ D {\bf 53}, 3474 (1996)
[arXiv:hep-th/9510053].
}

\lref\cederwall{
T.~Adawi, M.~Cederwall, U.~Gran, B.~E.~W.~Nilsson and B.~Razaznejad,
``Goldstone tensor modes,''
JHEP {\bf 9902}, 001 (1999)
[arXiv:hep-th/9811145].
}

\lref\chs{
C.~G.~.~Callan, J.~A.~Harvey and A.~Strominger,
``Worldbrane actions for string solitons,''
Nucl.\ Phys.\ B {\bf 367}, 60 (1991).
}

\lref\diacon{E. Diaconescu and D. Freed, unpublished.}

\Title{\vbox{\rightline{EFI-04-25}\rightline{QMUL-PH-04-03} } }
{\vbox{\centerline{The Self-Dual String and Anomalies in the
M5-brane}}}
\vskip10pt
\baselineskip=12pt

\centerline{David S. Berman\footnote{$^a$}{D.S.Berman@qmul.ac.uk}}
\medskip
\centerline{\sl Department of Physics}
\centerline{\sl Queen Mary College, University of London}
\centerline{\sl Mile End Road, London E1 4NS, England}

\medskip

\centerline{ {\it and}} 

\medskip
\centerline{Jeffrey A. Harvey\footnote{$^b$}{harvey@theory.uchicago.edu}}
\medskip
\centerline{\sl Enrico Fermi Institute and Department of Physics}
\centerline{\sl University of Chicago}
\centerline{\sl 5640 S. Ellis Ave., Chicago, IL 60637, USA}

\baselineskip=14pt

\vskip 2cm
\noindent
\centerline{\sl Abstract}

We study the anomalies of a charge $Q_2$ self-dual string solution in the 
Coulomb branch of $Q_5$ M5-branes. Cancellation of these anomalies
allows us to determine the anomaly of the zero-modes on the self-dual
string and their scaling with $Q_2$ and $Q_5$. The dimensional
reduction of  the five-brane anomalous couplings then lead to certain 
anomalous couplings for D-branes.

\Date{8/04}


\newsec{Introduction}

There remain many puzzling aspects concerning coincident branes in
M-theory. One of the central puzzles involves the lack of a microscopic
derivation of the number of degrees of freedom on $Q_5$ coincident five-branes.
Some information about the theory has been obtained through circuitous methods
such as anomalies \harv, low energy scattering \Klebanov\ and the
AdS/CFT correspondence \henningson. All these methods show that in the
large $Q_5$ limit the number of degrees of freedom scale as $Q_5^3$ for the
five-brane. For the two-brane the number of degrees of freedom scales
like $Q_2^{3/2}$.

In the five-brane itself there are self-dual strings \refs{\Callan,
\Gibbons,\howe}. These
appear as solitonic solutions to the nonlinear five-brane theory
and are associated with the five-brane worldvolume description of a 
membrane ending on a five-brane. From
the point of view of a spontaneously broken five-brane theory,  these
strings are similar to W-bosons, becoming tensionless when the
five-brane separation of the branes vanishes. There is no adequate
description of these self-dual strings. Apart from the problem of
describing tensionless string dynamics, self-dual strings are never
weakly coupled and so one cannot use standard perturbative
techniques. Furthermore, the five-brane theory in
which these strings live has no known description when there is more
than one five-brane.

This paper will be concerned with the study of these strings, in
particular, with $Q_2$ coincident self-dual strings. In \berman\  it
was  found that the absorption cross section of $Q_2$ coincident
self-dual strings (when the number of five-branes $Q_5=1$) is
proportional to $Q_2$ indicating that number of degrees of freedom
scales linearly with $Q_2$. A similar scattering calculation was done
for  a membrane probing a  supergravity solution describing $Q_5$
coincident five-branes,  indicating  that the number of degrees of
freedom scale as $Q_5$ (this is with $Q_2=1$).  (Note, the same
reasoning applied to D3 branes where the absorption cross section for
$Q$ D3 branes was shown to scale as $Q^2$ indicates a $Q^2$ scaling in
the number of light degrees of freedom as is expected for $Q$
coincident three-branes \Klebanov.) 

Here we wish to consider a more general situation and use anomalies to
determine the scaling of the number of degrees of freedom for $Q_2$
self-dual strings in a generic spontaneously broken five-brane theory
associated with $Q_5$ five-branes in the Coulomb branch. We will be interested 
in the dependence of the anomaly on the charges $Q_2$ and $Q_5$ as well as
the role of the unbroken gauge group.  The power of anomalies is that they
are topological in nature and can be studied in the low-energy theory, yet
provide a probe into high energy physics. This will prove very
powerful in this situation where a description of the fundamental
degrees of freedom is lacking and we have only an effective low energy
description.

We will consider a five-brane theory labelled by an ADE Lie algebra (how
this symmetry is realised in terms of local fields is not known (see
\sevrin\ for a discussion of this problem). The simplest case is the U(N) 
theory, which is actually a $U(1) \times
SU(N)$ theory with the decoupled U(1) corresponding to the Nambu-Goldstone
mode of translating the whole stack of branes. Spontaneous
symmetry breaking of the theory occurs when one (or more) of the
branes are separated off from the stack. There will then be a U(1)
mode  corresponding to fluctuations in the separation of the 
brane  stacks. At low energies, i.e. at scales less than the inverse brane
separation, that U(1) mode will be described by a U(1) (0,2) tensor
multiplet. 

Self-dual strings are solutions of the (0,2) abelian tensor
multiplet. As such we can embed the known self-dual string solution
into the U(1) tensor multiplet corresponding to this separation mode
as opposed to the usual overall U(1) translation mode. This will allow
us to investigate properties of the spontaneously broken five-brane
theory. In particular, we will be able to see how anomalies in the normal 
bundle of the self-dual string may be cancelled by inflow from Wess-Zumino type
terms in the five-brane world volume theory. Imposing this
cancellation will then allow us to determine the scaling of the
coefficient in terms of $Q_2$ and $Q_5$ and the dependence on the
unbroken  gauge group.  The inflow mechanism is analogous to the sort
used to cancel anomalies of intersecting D-branes
\refs{\michael,\Cheung,\Minasian}  or the M-theory five-brane itself, \refs{\harv,\wittfive}  with an inflow from the supergravity bulk. 

Now the {\it{bulk}} is the five-brane world volume and the defect is the
self-dual string. Similar issues concerning the anomalies of self-dual
strings have been discussed in \refs{\brax,\lechner}; here we expand their
discussion to multiple branes and relate the couplings involved in
anomaly cancellation to interactions discussed in the recent
literature \refs{\ganor,\intril}.
While this paper was in preparation we became aware of \Henningson\ which also
considers anomaly cancellation for strings in a six dimensional theory but does not consider the anomalies normal to the five-brane nor allow for a scaling of the zero modes with the charge. 

Apart from providing an insight into the self-dual string, the terms
that are required by anomaly inflow on the five-brane imply certain anomalous 
couplings for D-branes via dimensional reduction.
These terms are closely related to terms discussed elsewhere in the literature
\refs{\thompson,\tseytlin} and will be discussed in the penultimate section of the paper.

\newsec{The Self-dual String Fermion Zero Modes and Their Anomalies} 

\subsec{The Self-Dual String Solution}

The Bosonic field content of the 5+1 dimensional (0,2) tensor multiplet theory
consists of a two form field $b$ whose three form field strength $h$
obeys a self-duality constraint \fiveeom\ and five scalar fields, $\phi^i, \,
i=1..5$. The self-dual string \howe\
is a half BPS solution with the two form $b$ field excited along
with a single scalar field, denoted here as $\phi$. The solution is
given explicitly by:
\eqn\stringsol{ \phi(r) = \phi_0 + {2 Q\over r^2} \quad {{h}}_{01p}(r)=\pm {1 \over 4}
\partial_p \phi(r) \quad {{h}}_{mnp}(r) = \pm {1 \over 4} \epsilon_{mnpq}\partial_q
\phi(r)  \quad m,n,p,q=2,3,4,5\, . }
r is the radial coordinate of the space transverse to the string
ie. for a string lying along $x^1$, $r^2=x_2^2 + x_3^2 +x_4^2 +x_5^2$
and  $\epsilon_{mnpq}$ is the associated epsilon tensor of this
transverse space. The charge of the string is, $Q_2=\mp Q$.

This solution can be viewed as the worldvolume description of an M2 or
anti-M2 brane ending on the M5-brane. The field $\phi$ represents the
value of one of the coordinates transverse to the M5. With the
convention that $\phi$ increases from left to right, the solution with
the upper choice of $\pm$ sign and $Q>0$ corresponds to an M2 coming
in from the right and terminating on the $M5$ located at $\phi_0$. The
lower choice of sign with $Q>0$ is then an anti-M2 coming in from the
right while the upper choice of sign with $Q<0$ is an M2 coming
in from the left and the lower choice of sign with $Q<0$ is an anti-M2
coming in from the left.

\subsec{Fermion Zero Modes}

This self-dual string solution preserves eight supercharges (one half BPS with respect
to the five-brane and one quarter BPS with respect to M-theory). As usual for such BPS
objects, the fermion zero modes of the lowest charge solution
are generated by the broken supersymmetries, that is by
$\epsilon^{\alpha l}$ which satisfy:
\eqn\fmodes{ \epsilon^{\beta j}= \mp (\gamma^{01})_\alpha ^\beta (\gamma_{5'})_i^j 
\epsilon^{\alpha i} }
where $\alpha, \beta=1..4$ are (Weyl) spinor indices of Spin(1,5) and
$i,j=1..4$ are spinor indices of USp(4) the Spin cover of the SO(5)
R-symmetry group. The choice of plus or minus is associated with the choice of sign in the solution in \stringsol . We now wish to decompose these Fermionic zero modes
into representations of:
\eqn\rep{\rm{Spin}(1,1)\times Spin(4)_T \times Spin(4)_N \, .}

This decomposition is the Spin cover of the Lorentz group that is
preserved by the self-dual string solution. (The original $SO(1,5)
\times SO(5)$ group which is preserved by the five-brane becomes
broken by the self-dual string solution to $SO(1,1) \times SO(4)_T
\times SO(4)_N$). The subscripts T, N denote tangent and normal to the
five-brane world volume respectively. An eigenspinor of $\gamma^{01}$
will be a 1+1 chiral spinor and an eigenspinor of $\gamma_{5'}$ will
be a Weyl spinor of the $Spin(4)_N$. Importantly, the 6d Weyl spinors
that are also eigenspinors of  $\gamma^{01}$ are Weyl spinors of
$Spin(4)_T$. Putting these facts together implies the BPS self-dual string has (4,4) supersymmetry in 1+1 dimensions with the Fermions lying in the
following representations of \rep\ :
\eqn\one{(2,1,1,2)^{1/2} \oplus (1,2,2,1)^{-1/2} \, ,}
and the anti-BPS (negative sign in \stringsol\ ) string has the Fermions lying in:
\eqn\aone{(2,1,2,1)^{1/2} \oplus (1,2,1,2)^{-1/2} \, . }

The superscript labels the $SO(1,1)$ helicity with the numbers in
brackets labelling the $Spin(4)=SU(2) \times SU(2)$ representations.

For solutions with charge $|Q| \ge 1$ there will be $|Q|$ such
multiplets.  One can argue for this result in several ways. Since the
self-dual string solution is BPS, solutions with $|Q|>1$ can be
deformed into $|Q|$ separated string solutions, each with its own
centre of mass Bosonic zero modes. Supersymmetry then requires that
each of these Bosonic multiplets be accompanied by fermion zero
modes. Alternatively, we can reduce this system to a D-brane
configuration (as will be discussed in detail later) and then the
$|Q|$ fermion zero modes arise from the usual Chan-Paton factors. In
principle it should also be possible to show this by analysing an
index theorem for the corresponding Dirac operator on the brane, but
to our knowledge this has not been done in the literature. (It would
be interesting to see this explicitly since the Dirac operator on a
brane in the presence of a background field is somewhat different from
the usual Dirac operator).

The above analysis has all been concerned with $Q_5=1$, that is
membranes ending on a single M5-brane. For $Q_5>1$ the zero mode
structure on strings corresponding to membranes ending on the stack
of fivebranes is unknown. The main result of this paper will be to put
constraints on the zero mode structure for this case.

\subsec{Zero Mode Anomalies}

We now turn to a computation of the anomaly of the fermion zero modes for
$Q_5=1$. We are interested in anomalies in diffeomorphisms of the eleven-dimensional spacetime 
which preserved the self-dual string solution of the M5-brane, or equivalently which
preserve the configuration of M2-branes ending on the M5-brane. 
These diffeomorphisms act as diffeomorphisms of the string world-sheet, or as
gauge transformations of the $SO(4)_T \times SO(4)_N$ normal bundle to the string.
Since there are equal numbers of left and right-movers, there is no anomaly in world-sheet
diffeomorphisms. 
However, the left and right moving fermions are in
different representations of the normal bundle (the R-symmetry group)
which will give rise to a normal bundle anomaly. (In a field theory context these would be the 't Hooft anomalies).

The anomaly can be computed by treating the each $SO(4)$ symmetry
as a gauge symmetry (see e.g.the discussion in \wittfive). In two 
dimensions the anomaly is derived by descent from a 
four-form characteristic
class. $SO(4)$ has two such classes, the first Pontryagin class and the
Euler class and the anomaly in this case is proportional to the Euler class.
For an $SO(4)$ field strength two-form $F^{ab}$, a,b,=1,2,3,4 the Euler class
is
\eqn\two{\chi(F) = {1 \over 32 \pi^2} \epsilon^{abcd}F^{ab} \wedge F^{cd}}
which in terms of $SU(2)$ field strengths is,
\eqn\three{{1 \over 4 \pi^2} (tr F_+^2 - tr F_-^2) \quad . } 
The descent procedure involves writing 
\eqn\four{\chi(F) = d \chi_3^{(0)}(A)}
and the gauge variation as
\eqn\five{ \delta \chi_3^{(0)} = d \chi_2^{(1)}(A) \, .}
The normal bundle anomaly for each SO(4) is then proportional to 
\eqn\six{ \int_{\Sigma_2}  \pi \chi^{(1)}(A)}
with $\Sigma_2$ the self-dual string world-volume. (The factor of $\pi$ appears instead of the usual $2 \pi$ since the Fermions are Majorana and Weyl.) The total anomaly
is
\eqn\anomaly{\int_{\Sigma_2} \pi  Q_2 (\chi_2^{(1)}(A_N) \mp \chi_2^{(1)}(A_T))}
with the sign correlated with the sign in \stringsol. From now on we work with
the upper sign in \stringsol\ and \anomaly.

In the next section we will consider the self-dual strings in a (0,2)
multiplet arising from the low energy description of fivebranes in the
Coulomb branch. We will show how the anomaly may cancelled by an inflow 
mechanism and in doing so we will also see how the anomaly must scale 
with the number of fivebranes.

\newsec{Anomaly Cancellation}

\subsec{Generalities}

There are various approaches and levels of analysis one can take in dealing
with anomaly cancellation in string theory and M-theory, particularly
in analysing anomalies for extended objects.

As usual in trying to cancel anomalies, one is free to add local counterterms
to the Lagrangian. In theories with UV divergences these can be viewed as
part of the definition of the theory. In theories such as string theory, and
presumably M-theory, such counterterms should in principle be computable
in the underlying microscopic theory. Thus, in the original analysis of Green
and Schwarz \gsanom, anomaly cancellation was understood both by a direct
string calculation, and in the low-energy effective Lagrangian, and the
local counterterms needed to cancel the anomaly in the low-energy effective theory 
could be verified directly. Similarly, the analysis of anomaly cancellation on
D-branes in string theory requires certain anomalous, or Wess-Zumino, terms in
the low-energy effective action on the D-brane, and it is possible to
check that the required terms are present by a direct calculation \CrapsI .

The situation for NS 5-branes and M5-branes in IIA string theory and M-theory
is less satisfactory. It was shown in \wittfive\ that the anomalies cancel for
NS 5-branes after addition of a local counterterm to the 5-brane low-energy
effective action, but to our knowledge this counterterm has not been verified
by a direct calculation in string theory. Similarly, anomalies cancel for the
M5-brane after one gives a careful definition of the $C$ field and its action
in the presence of an M5-brane \freed. This definition involves adding local
counterterms to the M-theory low-energy effective action. There is however
no microscopic derivation of these terms. However,the
counterterm required in \wittfive\ does follow from the analysis of \freed, so
at least they are not independent problems \beckers. 

In dealing with anomalies on extended objects there are two different
points of view one can take. The NS 5-brane and M5-brane can be viewed
as smooth soliton solutions of string theory or M-theory. Viewed this way,
the zero modes localised on the 5-brane arise from a collective coordinate
expansion of the bulk fields  \refs{\chs,\kaplan,\cederwall}. However, it has proved difficult
to analyse anomalies directly in this framework because this involves
dealing with the bulk Rarita-Schwinger equation in the
non-trivial fivebrane geometry. The point of view which is most commonly 
taken is to split the degrees of freedom into bulk degrees of
freedom and degrees of freedom localised on the brane without taking into
account the detailed form of the brane soliton solution or the relation between
localised zero modes and bulk fields. This is the point of view adopted
in \freed.

In this paper we will take this last point of view. As we discussed earlier,
the M2-brane ending on an M5-brane can be viewed as a soliton solution to the
M5-brane equations of motion, but for analysing anomalies we will treat the
M2-brane and M5-brane zero modes as independent of the bulk fields.
We also allow ourselves the freedom to add local counterterms to the Lagrangian
as long as they respect the symmetries of the system. As we discussed earlier,
this means they should respect the $Spin(1,1) \times SO(4)_T \times SO(4)_N$
symmetry of the M2-M5 configuration.

\subsec{Anomaly Inflow for $Q_5=1$}

The cancellation of the $SO(4)_T$ anomaly in \anomaly\ is the most
straightforward to understand and was already discussed in \brax. We redo
this analysis here using the formalism developed in 
\refs{\freed, \beckers}\foot{In the conventions
of \brax\ we have chosen units with $T_3=1$, $\kappa_{11}=1/T_6=\pi$.}.

We denote the M5-brane and M2-brane world-volumes by $\Sigma_6$ and $\Sigma_3$
respectively. The M2-brane boundary on the M5-brane is the worldvolume
of the self-dual string with worldvolume $\Sigma_2 \equiv \partial \Sigma_3$.
We introduce two bump forms $d \rho (r)$ and $d \rho'(r')$ where $r$ is the 
radial direction transverse to $\Sigma_2$
in $\Sigma_6$ and $r'$ is 
the radial direction transverse to $\Sigma_6$ in the 11-dimensional spacetime
manifold $M_{11}$.  M-theory in $M_{11}$ has, in the absence of M5-branes, a
four-form field strength $G_4$ with $G_4= dC_3$ and a Bianchi identity
$dG_4=0$. As we mentioned earlier, the M5-brane worldvolume theory has, in
the absence of self-dual strings, a three-form field strength $h_3$ with
a Bianchi identity $d h_3 = - G_4|_{\Sigma_6}$. In the presence of M5-branes and 
self-dual strings the resulting Bianchi 
identities become
\eqn\bian{\eqalign{d G_4 & = 2 \pi \delta_5(\Sigma_6 \hookrightarrow M_{11}) \cr
                   d h_3 & =  \pi Q_2 \delta_4(\Sigma_2 \hookrightarrow \Sigma_6)
                             - G_4|_{\Sigma_6} }}

Note the factor of $\pi$ as opposed to $2 \pi$ on the right hand side for the Bianchi identity of $h_3$. This arises because the flux quantisation for Dyonic strings in six dimensions is given by $eg=\pi n$ \pope .
Physically, the quantities $\delta_5$ and $\delta_4$ in \bian\ are often thought of
as delta functions with integral one in the spaces transverse to $\Sigma_6$
and $\Sigma_2$ respectively. However, a more careful mathematical treatment
has turned out to be necessary in dealing with anomalies in the M5 system in 
which we think if these as the Poincare duals to $\Sigma_6$ in $M_{11}$ and
$\Sigma_2$ in $\Sigma_6$ respectively. Using the isomorphism between the Poincare
dual and the Thom class of the normal bundle we can choose explicit representatives
for $\delta_5$ and $\delta_4$ given by
\eqn\delforms{\eqalign{ \delta_5(\Sigma_6 \hookrightarrow M_{11}) & = d \rho'(r')
\wedge e_4/2 \cr
                        \delta_4(\Sigma_2 \hookrightarrow \Sigma_6) & = d \rho(r) \wedge
e_3/2 \cr }}
where $e_4$ is the global angular form with integral two over the $S^4$ fibres
transverse to $\Sigma_6$ in $M_{11}$, and $e_3$ is the global angular form with
integral two over the $S^3$ fibres transverse to $\Sigma_2$ in $\Sigma_6$.

Following \refs{\freed,\beckers} we write
\eqn\splite{e_3/2 = d \psi_2 + \Omega_3, }
with $d \Omega_3 = - \chi(F_T)$ and solve these Bianchi identities so that
$G_4$ and $h_3$ are non-singular on $\Sigma_6$ and $\Sigma_2$ respectively:
\eqn\bsolve{\eqalign{G_4 & = d C_3 - 2 \pi d \rho' \wedge e_3^{(0)}/2 \cr
                     h_3 & = d b_2 - C_3 +  \pi Q_2
                     (\rho \Omega_3 - d \rho \wedge \psi_2). \cr }}
Since $h_3$ must be gauge invariant, and the variation of $\Omega_3$
under $SO(4)_T$ gauge variations is $\delta \Omega_3 = d \chi_2^{(1)}(A_T)$,
we learn using $\rho(0)=-1$ that $b_2$ has a $SO(4)_T$ gauge variation so
that the minimal coupling of the string to the two-form field on the
M5-brane has a variation
\eqn\bvary{\int_{\Sigma_2} \delta b_2 =  \pi Q_2 \int_{\Sigma_2} \chi_2^{(1)} }
which cancels the $SO(4)_T$ anomaly in \anomaly. 

Where we differ from \brax\ is in the treatment of the $SO(4)_N$ anomaly.
In the above formalism $G_4|_{\Sigma_6} = d C_3|_{\Sigma_6}$ and there
does not seem to be room for the additional $SO(4)_N$ variation of $b_2$
found in \brax.

Fortunately, the $SO(4)_N$ anomaly can be cancelled by adding a local counterterm
to the M5-brane action:
\eqn\fcounter{S_N = - \int_{\Sigma_6} h_3 \wedge \chi^{(0)}(A_N) }
It is easy to see using the second equation in \bian\ that this term has an anomalous
variation localised on $\Sigma_2$ that cancels the anomaly in \anomaly. In
the following section we will see that a coupling of the form \fcounter\ is already
know to exist in the theory of $Q_5>1$ M5-branes.  

We have thus demonstrated anomaly cancellation for $Q_2$ M2-branes ending on
a single M5-brane. For $Q_5>1$ M5-branes we do not know how to carry out such
a general analysis, because neither the fermion zero mode structure nor the
$Q_5$ dependence of the coupling \fcounter\ is known. However, we can obtain some
partial results by utilising a known generalisation of the coupling \fcounter\
which appears on the Coulomb branch of the theory with $Q_5>1$.

\subsec{Anomaly Inflow for $Q_5>1$}

The analysis in the previous section involved coupling to the $U(1)$ centre of
mass multiplet of the $(2,0)$ theory. We now consider a general M5-brane theory
with an ADE symmetry $G \times U(1)$ broken down to $(H \times U(1)) \times U(1)$.

As in the analogous D1-D3 system, it is expected that this theory will have self-dual
strings which couple to the $U(1) \subset G$ factor. These are the analogs of the
non-Abelian 't Hooft-Polyakov monopoles in the D1-D3 system. Assuming that such 
solutions exist and act as sources for the $h$ field in the relative $U(1)$
multiplet, we can deduce some facts about the anomalies of the fermion zero
modes on such strings. 

The coupling in the five-brane theory which is anomalous in such a string
background 
originates from a coupling derived in
\refs{\ganor,\intril}.
Following the conventions in \intril\ the
coupling is given by
\eqn\seven{S_N=\alpha_e \int_{\Sigma_6} h_3 \wedge \Omega_3(\hat \phi, A)}
where
\eqn\chern{\eqalign{d \Omega_3 = &{1 \over {64 \pi^2}} 
\epsilon_{a_1..a_5}[(D\hat\phi)^{a_1} \wedge (D \hat\phi)^{a_2} \wedge 
(D \hat\phi)^{a_3} \wedge (D \hat\phi)^{a_4} \cr -&2 F^{a_1 a_2} \wedge (D \hat\phi)^{a_3} \wedge (D \hat\phi)^{a_4}  +
F^{a_1 a_2} \wedge F^{a_3 a_4} ] \hat\phi^{a_5}  }}  with  $(D_i \phi)^a = \partial_i \phi^a -
A_i^{ab} \phi^b$ the covariant derivative of $\phi^a$ with $A_i^{ab}$
the SO(5) gauge field of the five-brane normal bundle and $\hat\phi^a=
{{\phi^a} \over{\| \phi \|}} $. The coupling constant $\alpha_e$
depends on the breaking  of the five-brane theory gauge group. If $G$ is the ADE Lie
algebra labelling the $(0,2)$ theory, then for a
breaking given by $G \rightarrow H \times U(1)$, it was argued that
\eqn\alphae{ \alpha_e= {1 \over 4 }( |G| -|H| -1 ) \, . } 
For example if $G= SU(Q_5+1)$ and $H= SU(Q_5)$  then, $\alpha_e ={1
\over 2}  Q_5$.

This coupling was derived by considering anomalies in the Coulomb
branch  of the five-brane theory, ie. when the scalars have non-zero
vacuum expectation values. The point is that the U(1) multiplet at low 
energies must include an interaction term to compensate for what would 
otherwise be a difference in the anomaly at the origin of moduli space and 
the anomaly at a generic point in the Coulomb branch. Hence the decoupling of 
the U(1) multiplet never really happens; this term is always sensitive to the 
full theory and so even in the infrared there remains some information of the 
integrated out ultra-massive modes.

Here we will evaluate this interaction term in the presence of a
self-dual string embedded in the naively decoupled U(1) tensor multiplet.

For the self-dual string solution we can take just
one scalar, say $\phi^5$, to be non-zero and obeying \stringsol\ ,
while also taking $A^{5a}=0$ to reduce the $SO(5)$ to the $SO(4)$
preserved by the string solution. Then it is easy to see that in the
presence of the self-dual string $\Omega_3 $ reduces to
$\chi^{(0)}(A)/2$. The effective coupling on the fivebrane in the
presence of the self-dual string is thus
\eqn\eight{S_{N}=- {\alpha_e \over 2} \int_{\Sigma_6} h_3 \wedge \chi^{(0)}(A_N) \, .}

In the presence of the self-dual string the coupling \eight\ is not gauge 
invariant. 
Rather,
its gauge variation is \eqn\nine{\delta S_{N} = - {\alpha_e \over 2}
\int_{\Sigma_6} h_3 \wedge d \chi^{(1)} = -{\alpha_e \over 2} \int_{\Sigma_6}
dh_3 \wedge \chi^{(1)} \, . } Now the self-dual string acts as a
source of $h_3$ via the equation
\eqn\ten{d h_3 =  \pi Q_2 \delta_4 (\Sigma_2 \hookrightarrow \Sigma_6) } 
Using
this, the variation \nine\ becomes
\eqn\eleven{- {{1\over 2} \pi \alpha_e } Q_2 \int_{\Sigma_2} \chi^{(1)} \, . }

Now, assuming the anomalous variation is, as before, cancelled by the fermion
zero modes we deduce that the $SO(4)_N$ zero mode anomaly of the
self-dual string in the Coulomb branch scales as:
\eqn\twelve{c={1 \over 2}  \alpha_e Q_2 \, .}

Using the above and the equation for $\alpha_e$, \alphae\ we may
compute the anomaly for the string, that is $c$ in terms of the
charges $Q_2,Q_5$.

Consider the case of pulling off a single five brane from a stack of
$Q_5+1$ fivebranes and  embedding  the string in the relative
U(1). The five-brane theory would have  $G=SU(Q_5+1)$ and $H=SU(Q_5)$
giving, $\alpha_e ={1 \over 2} Q_5$. This would  then give
\eqn\c{ c ={1 \over 4} Q_2 Q_5 \, .} 

We may then interpret $c$ as being related to the number of degrees
of freedom of the string. The $Q_2 Q_5$ dependence of the anomaly is
then  consistent with the cross section scattering calculation of the
self-dual  string described in \berman .

A more interesting situation arises if one considers a different
breaking pattern for the five-branes. Take a stack and separate all
the branes thus giving a maximal breaking with $G=SU(Q_5+1)$ and
$H=U(1)^{Q_5}$. In this case simply applying the above formula yields,
$ c= {1 \over 8} Q_2 (Q_5^2 +Q_5-1)$. In this case
no cross section scattering
calculation has been done to confirm the charge dependence. It would be
interesting to find other ways to study this system which would confirm
this behaviour.

Note that for this calculation we are really only using the
supersymmetry/R-symmetry preserved by the string solution to determine
the anomaly in conjunction with  the Ganor, Motl, Intrilligator term
\chern\ for the cancellation and so the precise form of the solution
should not matter. Thus even if one might be concerned about applying
the solution of \howe\ in the more exotic circumstances
described above, provided the symmetries of the string are the same our 
results should remain valid.

\newsec{Dimensional Reduction}

We now consider the implications of these terms for IIA string theory
by reducing M theory on an $S^1$. To be explicit we take the M5-worldvolume
to lie in the $(0,1, \cdots 5)$ plane and the M2-worldvolume to lie in the $(016)$
plane. The self-dual string worldvolume then lies in the $(01)$ plane. 
Normal bundle gauge transformations that preserve this configuration act
act on the $(2,3,4,5)$ coordinates ($SO(4)_T$) or the $(7,8,9,10)$ coordinates
($SO(4)_N$). 

There are then two interesting reductions to IIA string theory. We can take
one of the $(2,3,4,5)$ coordinates to be periodic. This turns the M2 into a
D2-brane in IIA theory and the M5 which wraps this periodic coordinate into
a D4-brane. The $SO(4)_T$ symmetry is broken to $SO(3)$ which has no Euler
class, so there is no $SO(3)$ normal bundle anomaly. However, the $SO(4)_N$
symmetry is preserved and has an anomaly derived by descent from the Euler
class. 

Cancellation of this normal bundle anomaly for a 
D2 ending on a D4 requires a coupling on the D4-worldvolume of the form
\eqn\deefour{\int_{\Sigma^5} F_2 \wedge \chi^{(0)}(A_N)}
where $F_2$ is the $U(1)$ gauge field strength on the D4-brane.
Note that this term is distinct from the usual anomalous couplings
on D-branes \refs{\michael,\Cheung,\Minasian}. Because it is independent of the
bulk Ramond-Ramond fields, it would arise at one-loop level rather
than as a tree-level coupling, presumably explaining why it has not been
seen in previous explicit calculations of anomalous couplings \refs{\Bershadsky,\CrapsI,\Morales,\CrapsII}.
In fact, closely related couplings (but for relative $U(1)$ factors) have been
discussed in \thompson\ and \tseytlin.

The other inequivalent reduction takes one of the $(7,8,9,10)$ coordinates
to be periodic. This turns the M2 into a D2 while converting the M5 into a
NS5-brane. Now the $SO(4)_N$ symmetry is broken to $SO(3)$ with vanishing anomaly
while the $SO(4)_T$ symmetry is preserved. Cancellation of this normal bundle
anomaly then requires a coupling on the NS5-brane worldvolume analogous to \eight .

\newsec{Acknowledgements}

We would like to thank C.~Bachas, N.~Copland, D.~Kutasov, M.~Green, P.~Howe, 
G. ~Moore, G.~Papadopolous, M.~Perry and A.~Ritz for relevant discussions. 
JH thanks D. Freed for discussions of his unpublished work with E. Diaconescu
on related material \diacon. 
DSB is supported by EPSRC grant GR/R75373/02
and would like to thank DAMTP and Clare Hall college Cambridge for continued 
support and CERN for hospitality during the final stage of the project. JH is supported by NSF grant PHY-0204608 and acknowledges the stimulating
atmosphere provided by the Aspen
Center for Physics during the completion of this work. 

\listrefs
\bye